\documentclass{article}

     \PassOptionsToPackage{numbers, compress}{natbib}

\usepackage[final]{neurips_2019}

\usepackage[utf8]{inputenc} 
\usepackage[T1]{fontenc}    
\usepackage{hyperref}       
\usepackage{url}            
\usepackage{booktabs}       
\usepackage{amsfonts}       
\usepackage{nicefrac}       
\usepackage{microtype}      
\usepackage{mathtools}
\usepackage[normalem]{ulem} 
\usepackage{bm}

\usepackage{graphbox}

\usepackage{xcolor}

\title{Increasing performance of electric vehicles in ride-hailing services using deep reinforcement learning}

\author{%
  Jacob F. Pettit, Ruben Glatt, Jonathan R. Donadee, Brenden K. Petersen \\
  Computational Engineering Division\\
  Lawrence Livermore National Laboratory\\
  \texttt{\{pettit8, glatt1, donadee1, petersen33\}@llnl.gov} \\
}

\begin{document}

\maketitle

\begin{abstract}
New forms of on-demand transportation such as ride-hailing and connected autonomous vehicles are proliferating, yet are a challenging use case for electric vehicles (EV).
This paper explores the feasibility of using \textit{deep reinforcement learning (DRL)} to optimize a driving and charging policy for a ride-hailing EV agent, with the goal of reducing costs and emissions while increasing transportation service provided.
We introduce a data-driven simulation of a ride-hailing EV agent that provides transportation service and charges energy at congested charging infrastructure.
We then formulate a test case for the sequential driving and charging decision making problem of the agent and apply DRL to optimize the agent's decision making policy.
We evaluate the performance against hand-written policies and show that our agent learns to act competitively without any prior knowledge.
\end{abstract}

\section{Introduction}

The transportation sector is responsible for a significant share of all CO\textsubscript{2} emissions from human activity, representing $37\%$ of emissions produced by California \cite{CARB_Scoping_17} and $29\%$ of U.S. emissions \cite{EPAEmissions}.
Looking toward the future, the U.S. Department of Energy research shows that the proliferation of new forms of mobility, such as connected and autonomous vehicles (CAVs), could result in a $200\%$ increase in energy consumption in the U.S. transportation sector by 2050 relative to their baseline \cite{DOEBounds}.
The state of California has identified transportation electrification in tandem with transitioning to carbon free sources of electric power generation as a key strategy for reducing emissions from the transportation sector \cite{CARB_Scoping_17}.

However, electric vehicles (EV) face many obstacles to reaching high adoption levels, especially for use in ride-hailing for transportation network companies (TNCs) such as Lyft and Uber or in future CAV based services.
EVs have less driving range and longer refueling times than internal combustion powered vehicles, as well as limited charging infrastructure.
A ride-hailing EV driver should consider many spatially and temporally varying factors when deciding when and where to charge. This includes time-varying electric energy prices and emissions, ride-hailing surge pricing, demand for transportation service, traffic conditions, and queues at charging stations.
Each of these items can exhibit strong seasonality with time of day, geographic patterns, and randomness, creating opportunities to optimize timing and location of charging.
An optimal driving and charging strategy (known as a \textit{policy}) could help ride-hailing EVs to charge at the times of day when electric energy costs and emissions are lowest, to reduce time spent waiting in queues, and maximize a driver’s revenues, reducing the economic barrier to adoption of EVs in ride-hailing and CAV fleets.
Such a policy could be implemented as a decision support tool for human drivers, or as part of a CAV fleet management system.
In this feasibility study, we focus on the case of optimizing the policy of a single agent as a first step, with its environment taken as exogenous and given.

\section{Approach \& Results}

To investigate the potential to optimize policies for operation of ride-hailing EVs, we developed a data-driven simulation of a ride-hailing EV driving for a TNC.
The simulation includes modules representing the TNC, EV, the network of EV charging stations, and the electric grid.
The TNC randomly generates requested trips depending on the time of day and the EV's current location. Trip information includes destination, distance driven, time elapsed while driving, and transportation revenue earned.
The probabilistic model of trips is estimated from the open source New York City taxicab dataset for the year 2015\footnote{Source ride data: \url{https://www1.nyc.gov/site/tlc/about/tlc-trip-record-data.page}}. In our simulation, we consider trips within Manhattan, and have discretized the map into a grid of 120 zones.
The EV tracks the battery state of charge, depleting the battery when driving or recharging the battery when charging, and has a battery capacity of 100kWh.
The EV charging station network defines the locations of charging stations and models exogenous EV use of charging stations with an assumed pattern for queuing wait times.
Charging stations are assumed to have a fixed charging power of 100kW.
The electric grid determines the energy prices and CO\textsubscript{2} emissions produced per energy unit charged, depending on the time of day.
California values are used for electric energy prices\footnote{Source commercial energy prices: \url{https://caltransit.org/cta/assets/File/Webinar Elements/WEBINAR-PGE Rate Design 11-20-18.pdf}} and electric grid emissions\footnote{Source emission data: \url{http://www.caiso.com/TodaysOutlook/Pages/Emissions.aspx}}.

The driving and charging decision making of a ride-hailing EV can be formulated as a Markov Decision Process (MDP) \cite{Puterman14}.
In the MDP framework, an agent receives some observation about the state of their environment, chooses an action, and receives some reward.
Solving an MDP is equivalent to finding the decision making policy that maximizes an expected reward over the lifetime of the agent.

Our simulation environment is based on OpenAI Gym \cite{brockman2016openai} with a discrete action space, where the EV agent chooses between charging the vehicle or accepting a new ride at every step.
The agent's observation space is a vector consisting of the EV battery level, time of day, expected battery usage if the agent were to perform a ride, expected charging cost if the agent were to charge, expected emissions produced by choosing to charge, expected time when the vehicle would finish charging, and the expected queuing duration at the nearest charging station.

The reward function is defined as
\begin{equation}
  \label{equRewardFunction}
  r =
    \begin{cases}
    -(c + \epsilon E) & \text{if choosing to charge} \\
    \xi & \text{if successfully provide a ride} \\
    -3(c + \epsilon E) & \text{if attempting to provide a ride with insufficient battery}
    \end{cases},
\end{equation}
where $c$ is the cost paid to charge the vehicle, $\epsilon$ is the emissions produced from charging the vehicle, $E$ is a positive coefficient used to alter how much the agent considers the emissions produced from charging the vehicle, and $\xi$ is revenue from completing a ride.
$E$ can be considered analagous to a carbon tax or carbon price in dollars per ton of CO\textsubscript{2} produced.

If the agent attempts to provide a ride with insufficient battery energy, it receives the reward ($-3(c + \epsilon E)$) and is forced to charge the vehicle. After the vehicle is charged, a new ride is generated and the episode continues.
This penalty term encourages the agent to learn to charge the vehicle before the battery is depleted.
Should the agent choose to charge the car, then it is assumed to drive to the nearest charging station and the car is charged until full.
Finally, if the agent successfully provides a ride, it drives the passenger to the dropoff point and is paid for providing transportation service.

Reinforcement Learning (RL) \cite{sutton2018reinforcement} approaches to solving MDPs have been extended in recent years to what are known as Deep Reinforcement Learning (DRL) algorithms \cite{arulkumaran2017deep}.
DRL has demonstrated superhuman performance on computer games \cite{MnihEtal15}, was used to beat the world's strongest players in Go \cite{silver2018general}, and has made progress towards knowledge sharing \cite{GlattEtal16} and multiagent learning \cite{SilvaEtal17}.
We believe that DRL is well suited to exploit the spatial and temporal arbitrage opportunities presented in the ride-hailing EV's driving and charging environment.

We applied the Trust Region Policy Optimization (TRPO) \cite{schulman2015trust} DRL algorithm, as implemented by OpenAI baselines \footnote{Algorithm implementation from \url{https://github.com/openai/baselines}}, to train an EV agent's policy for our environment.
The policy and value function networks are each two-layer perceptrons with 64 units per layer and hyperbolic tangent activation functions. Training batch size was set to 4096, and the discount factor was set to 0.8.
Each simulation episode consists of one week of simulated time, and we assume that the vehicle provides uninterrupted service for the duration of the week.

For comparison, we also created heuristic policies that select when to charge based solely on the current battery level. Given a threshold $\lambda$, the handcrafted policies decide to charge when the battery level falls below $\lambda$, otherwise they accept the ride. We found $\lambda = 10\%$ to yield the highest performance, and we included baselines for $\lambda = 25\%$ and $\lambda = 50\%$ for additional comparisons.

\begin{figure}[h]
  \centering
    (a)\includegraphics[align=c, width=2.2in]{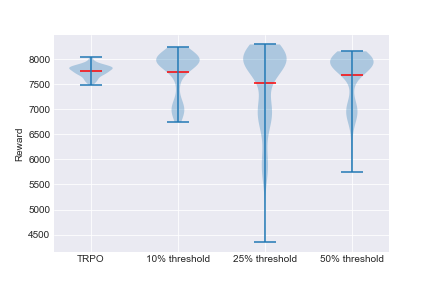}
    (b)\includegraphics[align=c, width=2.2in]{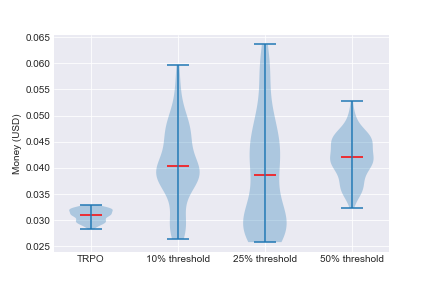}
    (c)\includegraphics[align=c, width=2.2in]{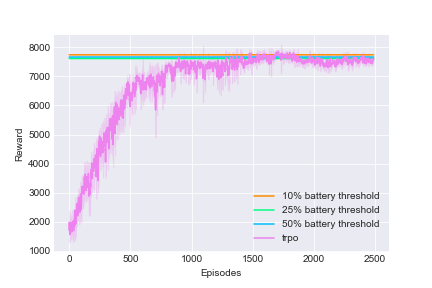}
    (d)\includegraphics[align=c, width=2.2in]{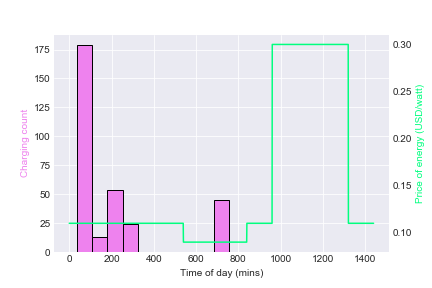}
    \caption{Experiment results: \textit{(a)} Reward earned by the trained policy (TRPO) in evaluation runs. \textit{(b)} Dollars spent on energy per mile driven in evaluation runs. \textit{(c)} Moving average of episodic rewards experienced by the agent over training. Error bar represents one standard deviation. \textit{(d)} Distribution of choosing to charge over the day peaks when energy prices are low.}
  \label{fig:Results}
\end{figure}

Our preliminary experiments show that our agent was able to achieve favorable results over the best performing heuristic policies.
In Figure~\ref{fig:Results} \textit{(a)} we see that the trained agent is able to achieve higher average reward than the heuristic policies while being much more consistent, which has a clear positive impact on dollar per miles as shown in Figure~\ref{fig:Results} \textit{(b)}.
The development of the reward during training is shown in Figure~\ref{fig:Results} \textit{(c)} where we see the average total episode reward over 2500 training episodes where each episode represents one week of driving.
Figure~\ref{fig:Results} \textit{(d)} shows the frequency of the agent's choice to charge by time of day.
The agent learns to charge most frequently in the middle of the night when energy prices and congestion at charging stations are low; or occassionally around noon, when people are at work, energy prices are at their lowest daily point, and wait times at charging stations are short.
The agent chooses not to charge at all during peak commute times, when surge pricing, charging wait times, and ride demand are greatest.

\section{Discussion}
Use of electric vehicles in existing and emerging forms of transportation, such as ride-hailing, public transit, and CAVs, presents numerous planning and operational challenges.
\citet{lin2018efficient} explore multi-agent DRL agorithms for optimizing operations of ride-hailing fleets, but without considering the many unique challenges related to EV charging.
\citet{Manoosh19} studys the centralized operation of a joint electric-grid and EV transportation system according to meso-scale average rates of demand for trips.
This approach doesn't address on-line operations in response to stochastic events.
\citet{chen2016operations} follow a heuristic procedure to identify location and number of charging stations, and then evaluate the performance of ride-hailing EV agents following threshold policies.
Our approach focuses on solving the immediate issues facing today's independent ride-hailing drivers that might adopt EVs, and intend to extend our work to future multi-agent CAV EV fleets.
We believe that DRL based approaches to optimizing the operation of EV fleets in transportation services remain an important research opportunity to be pursued.
With further work on design of agent's observations, policy neural network architecture, and hyper parameter selection, performance can be greatly increased beyond what we have demonstrated here.


\subsubsection*{Acknowledgments}

This work was performed under the auspices of the U.S. Department of Energy by Lawrence Livermore National Laboratory under contract DE-AC52-07NA27344.
Lawrence Livermore National Security, LLC.
LLNL-CONF-789379.

\bibliographystyle{plainnat}
\bibliography{main}

\end{document}